\documentstyle[prl,aps,multicol,epsf]{revtex}

\def\dend#1{{\if*#1{\it Paenibacillus dendritiformis}\else
                \if-#1{\it P. dendritiformis}\else
                 {\it P. dendritiformis} #1\fi\fi}}
\def\Tvar{var. {\it dendron}}
\def\Cvar{var. {\it chiralis}}
\def\Tname#1{{\if*#1\dend* \Tvar\else
                \if-#1\dend{}\Tvar\else
                 \dend{}\Tvar{} #1\fi\fi}}
\def\Cname#1{{\if*#1\dend* \Cvar\else
                \if-#1\dend{}\Cvar\else
                 \dend{}\Cvar{} #1\fi\fi}}
\def\bsub#1{{\if*#1{\it Bacillus subtilis}\else
                \if-#1{\it B. subtilis}\else {\it B. subtilis} #1\fi\fi}}
\def\bacil#1{\if *#1{Bacillus}\else{B.}\fi}

\def\T{{${\cal T }$} }

\def\C{{${\cal C }$} }
\def\Ce{{${\cal C }$}}

\def\Tme#1{{\T morphotype#1}}

\def\be{\begin{equation}}
\def\ee{\end{equation}}
\def\ben{\begin{enumerate}}
\def\een{\end{enumerate}}
\def\ba{\begin{eqnarray}}
\def\ea{\end{eqnarray}}

\def\etal{{\it et al. }}
\def\text#1{{\hbox{#1}}}

\newcommand{\R}{{\sf R\hspace*{-0.9ex}\rule{0.15ex}%
 {1.5ex}\hspace*{0.9ex}}}
\def\re{\text{Re}}
\def\im{\text{Im}}

\def\bd{b}

\begin{document}

\title{Orientation field model for chiral branching growth 
of bacterial colonies}
\author{Inon Cohen
 and Eshel Ben-Jacob}
\address{School of Physics and Astronomy, Raymond \& Beverly Sackler
Faculty of Exact Sciences, Tel-Aviv University, Tel Aviv 69978,
Israel}

\maketitle

\begin{abstract}
We present a new reaction-diffusion model for chiral branching growth of
colonies of the bacteria \dend*.
In our model the bacteria are represented by a density field with
non-linear diffusion and a complex scalar field which represents
bacterial orientation. The orientation field introduces anisotropy
into the flux of bacteria, representing self-propulsion along their long
axes.
The model can also reproduce tip-splitting growth of other strains
(shorter bacteria) of the same species.
The model can capture changes of small number of bacteria, thus it can
be used to study the open question of transitions between
tip-splitting and chiral dendritic growth.
\end{abstract}

\begin{multicols}{2}
\narrowtext

Bacteria display various chiral properties. Mendelson \etal
\cite{Mendelson_Chiral} showed that long cells of \bsub* can
grow in helices, in which the cells form long strings that twist
around each other. They have also showed that the chiral
characteristics affect the structure of the colony.
Matsuyama \etal \cite{Matsuyam_WeakChiral} have found that colonies of \bsub
can grow into tip-splitting patterns with a global rotation about the
center of the colonies (global twist). Ben-Jacob \etal
\cite{pre_CW,BCSCV95} have found that some strains of \dend*
(see Ref.  \cite{TBG99} for identification of the bacteria) exhibit
similar patterns with global twist. They have also found that other
strains (referred to as \C (chiral) morphotype) present
different chiral property, chiral dendritic growth with local twist
(which they refer to as strong chirality).

\begin{figure}[htbp]
\centerline{
 a
 \epsfxsize=1.5in
 \epsffile{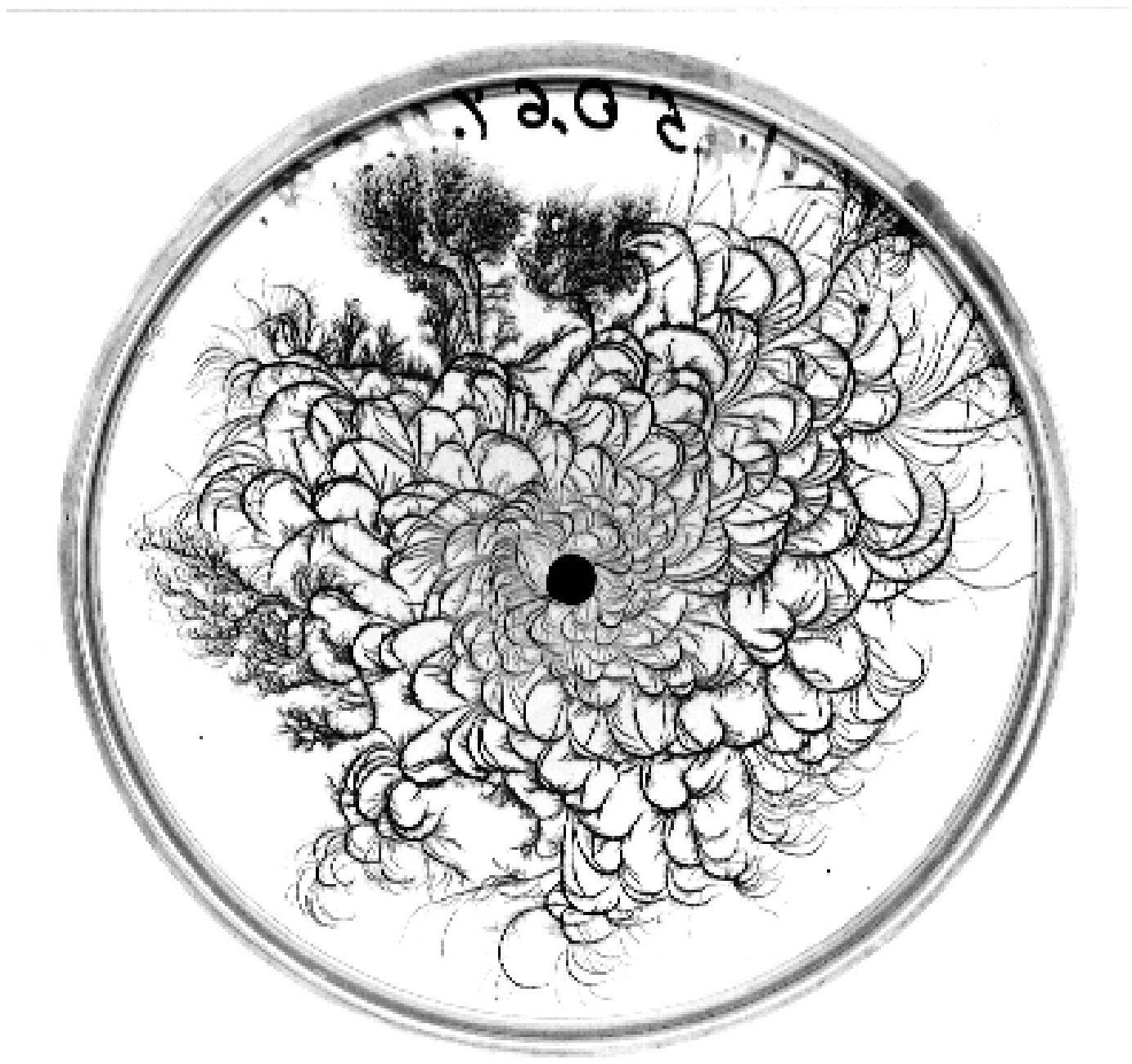}
 \ \ \ b
 \epsfxsize=1.3in
 \epsffile{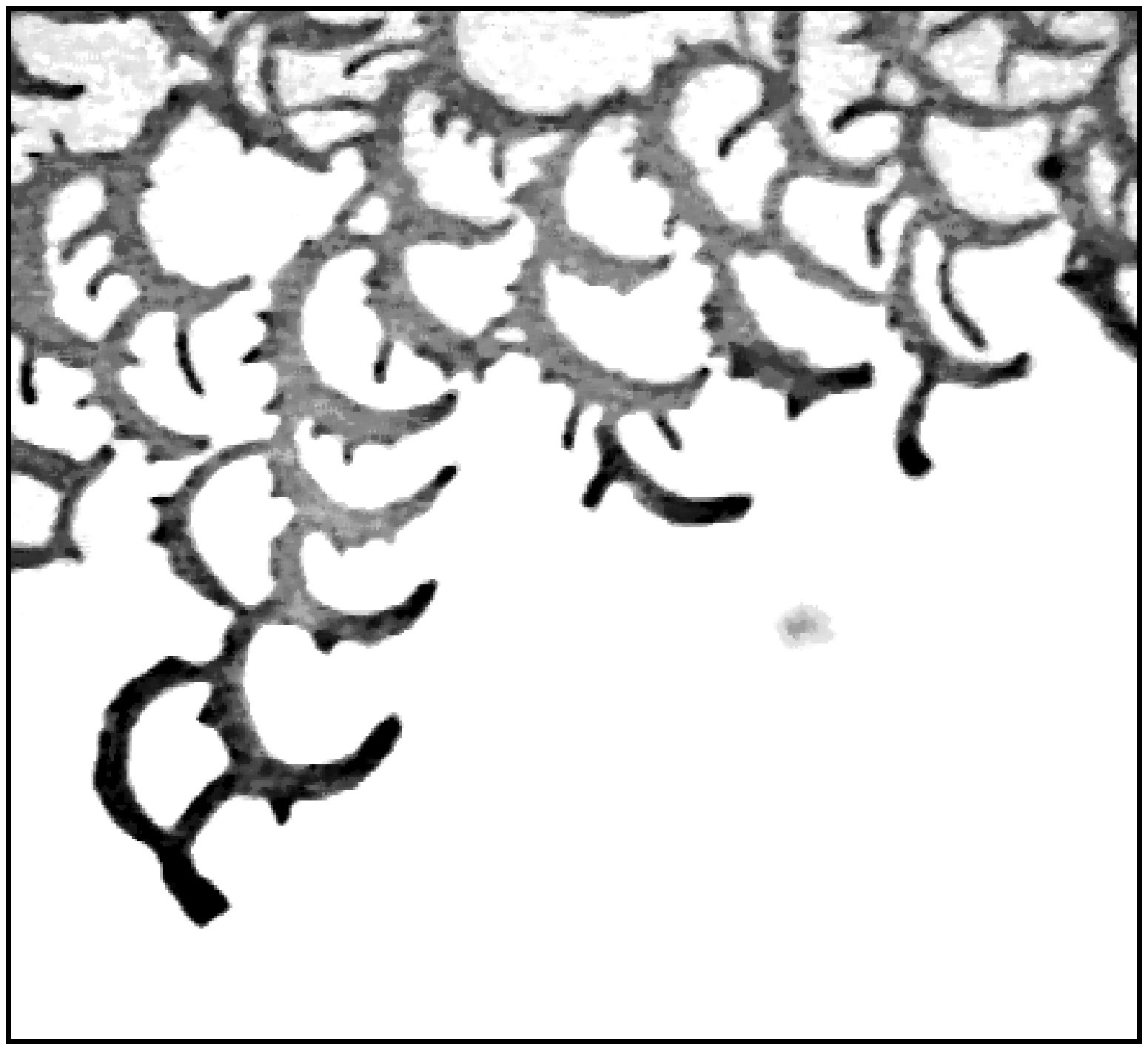}
}
\caption {
\label{fig:exper}
Chiral colonial growth of \dend*, strain \Ce.
a) Global view of a colony shows thin branches, all twisted with the
same handedness. Colony is grown at 2$g/l$ peptone level and 1.25\% agar
concentration.
b) Optical microscope observations of branches of \C colony.
$\times20$ magnification of a colony at 1.6$g/l$ peptone level and
0.75\% agar concentration, the anti-clockwise twist of the thin
branches is apparent. The curvature of the branches is the same
throughout the growth.
}
\end{figure}

Colonial patterns of \C bacteria grown on semi-solid agar are
characterized by chiral dendritic branching patterns, where the
branches are narrow and are twisted with the same handedness (Fig.
\ref{fig:exper}).
All colonies of this bacteria grown in similar conditions show the
same handedness.
Side branches are usually emitted to the convex side of the arced
branches (see Refs.
\cite{pre_CW,BCSCV95,BCGK99} for morphologies and studies of \C
bacteria).

2D chiral branching patterns are also observed in non-living systems.
Shapes resembling Sea Horse, or S, are formed during deposition of
thin films of fullerene-tetracyanoquinodimethane (C60-TCNQ) or pure
TCNQ \cite{SeaHorse_All}. Modeling of the system indicates that the
apparent curvature of the branches is actually a strong bias in
selection of splitting branches, and the branches themselves are not
curved.
The patterns of bacterial colonies share more resemblance with
patterns formed during compression of monolayers of various chiral
molecules at the air-water interface \cite{ChiralLipids_all}, and
electro-chemical deposition under a uniform magnetic field \cite{CHL99}.
Various modeling attempts indicates that the processes forming these
patterns are related to processes of solidification (see Refs.
\cite{our_reviews,GKCB98} for the differences between the
processes of solidification {\it vs.} colonial growth).
A different model is required for chiral branching growth of bacterial
colonies.

The Communicating Spinors model of Ben-Jacob \etal
\cite{BCSCV95,BCGK99} is an atomistic (discrete entities) model which
describes chiral branching growth of bacterial colonies.
The model presented here is not a mean field model of the
spinors model. The two models complement each other and
highlight different biological features.
Each spinor in the spinors model represents a large groups of about
$10^2$-$10^4$ bacteria.
This coarse graining  makes simulations
computationally feasible, but prevents modeling of processes
which cannot be averaged over large groups (specifically,
single-cell events). Continuous model is more appropriate
for this purpose (when interpreting densities as {\em
probability densities}) \cite{GCB99}. The
model presented here can be used to study the effect
of mutations and transitions on colonial morphologies, as will be
shown elsewhere.

Detailed description of the materials and methods used in the
experiments can be found in Ref. \cite{BCGGTHR00}.
The bacterial colonies are grown 
on top of agar (semi-solid  jelly) with peptone as a nutrient.
During colonial growth bacteria are confined to within a layer of
lubricating fluid, which is extracted from the agar by the bacteria
themselves. 
Inside the layer the bacteria
propel themselves in a random-walk like motion.
Optical microscope observations indicate that the length of \C
bacteria is up to 50 times their width.
In narrow branches, bacteria are aligned with their neighbors,
and their movement is quasi 1D random walk along their long axis.


In order to capture the details of bacterial dynamics we
define the
bacterial density per angle $ \bd(\vec{x},\theta ,t)$,
where $\vec{x} \in \R^2$ is a position and
$\theta \in [0,\pi ]$ is an angle of orientation.
We also use the bacterial density $B(\vec{x},t)$,
which is defined as the mean of $\bd$ over $\theta$.
The equation for the bacterial dynamics is
\begin{equation}
\dot{\bd}=\nabla \cdot \left[D_{b}(B) {\bf D_0}(\theta )\nabla \bd
\right] -\partial_\theta J_\theta +G\!\left( \bd , n\right)
\label{eq:b:0}
\end{equation}
where $\nabla$ operates in the spatial dimensions, $D_{b}$ is a
non-constant diffusion coefficient, $J_\theta$ is the flux in the
angle dimension which represents changes in bacterial orientation, $G$
is a reaction term representing growth and death, and $n$ is the
nutrient concentration.
We take the diffusion matrix 
${\bf D_0}\left( \theta \right) = {\bf R}(\theta ) ^T\left( 
\begin{array}{cc}
D_d & 0 \\ 
0 & D_l
\end{array}
\right) {\bf R}(\theta ) $,
where ${\bf R}(\theta ) $ is a rotation
matrix. $D_d$ is a constant coefficient
for diffusion in the bacterial direction $\theta$ (due to
self-propulsion forward and backward) and $D_l \le D_d$ is a constant
coefficient for diffusion in the lateral direction, due to
fluctuations.
Eq.  (\ref{eq:b:0}) is invariant under rotation in space and translation
in $ \theta $ by the same angle.

It was shown \cite{GKCB98,Kitsunezaki97,Cohen97,KCGB99,CGKB99} that
bacterial movement in a self-produced layer of fluid can be
approximated by a non-linear diffusion, where the diffusion coefficient
is proportional to the bacterial density to a power
greater than one. The proportion constant is related to the agar
dryness through the rate of absorption of the fluid into the agar.
Hence we take $D_{b}(B)=B^k$ ($k\ge2$), where
the proportion constant is included in $D_d$ and
$D_l$.  We will assume that bacterial friction with the agar is
proportional to the agar
dryness, and as the bacterial velocity is inversely proportional to the
friction, $D_d$ and $D_l$ are proportional to
the agar dryness to the power $-2$.

Following \cite{Kitsunezaki97,GKCB98} we take a simple form for the
growth function,
$G\!\left(\bd,n\right)=\bd (n - \mu)$, where $\mu \ge 0$ is a rate of
conversion into immobile sporulating cells. A more accurate form
should have saturation for high values of $n/B$, but the linear form
is a reasonable approximation for low
levels of nutrients, as is the case in the bacterial colonies.

Bacteria change their orientation ($J_\theta$) in response to their
neighbors orientation. It is  convenient to use an
auxiliary complex orientation field $p$, which is
defined as
$ p(\vec{x},t)= \frac 1{\pi }\int_0^{\pi }\bd(\vec{x},\theta,t)
e^{-i2\theta }\,d\theta$.
The mean orientation of the bacteria is along the vector 
$\left( \pm\re\sqrt{p}, \mp\im\sqrt{p} \right)$.

We found that it is sufficient to track the dynamic of $B$ and
$p$, instead of the full dynamic of $\bd$.
Using Fourier expansion of Eq.  (\ref{eq:b:0}) in $\theta $ and taking
the first two resulting equations we derive the Orientation Field (OF)
model, which includes equations for $B$, $p$, the nutrient
concentration $n$ 
and the density of immobile (sporulating) bacteria $s$
(at this stage $s$
is used only to record the history of the colonial development):
\begin{eqnarray}
\dot{B} &=&\nabla \cdot \left\{ B^k\left[ D_1\nabla B+2\,{\hbox{Re}}
\!\left( {\bf D_2} \nabla p\right) \right] \right\} +B(n-\mu ) 
\label{eq:OF:B} \\
\dot{p} &=&\nabla \cdot \left\{ B^k\left[ {\bf D_2^{*}}\nabla B+D_1\nabla
p\right] \right\} +p(n-\mu )
\nonumber \\ && 
+a\!\left( B,|p|\right) p+\gamma \!\left(
B^{k-2}\nabla B,p\right) ip 
\label{eq:OF:p} \\
\dot{n} &=& D_n \nabla ^2n-nB 
\label{eq:OF:n} \\
\dot{s} &=&\mu B
\label{eq:OF:s}
\end{eqnarray}
where $n$ is scaled to units of bacterial mass and 
$D_n$ is its diffusion coefficient.
$a$ and $\gamma$ are real valued functions, a decomposition into
orthogonal elements of the derivation of 
$\partial_\theta J_\theta$
(we give here the functions $a$ and $\gamma$, not the original $J_\theta$).
Here $D_1=\frac{ D_d+D_l}{ 2}$ and 
${\bf D_2} =\frac{D_d-D_l}{4}{\bf M}$ , where
${\bf M} =\left(
\begin{array}{cc}
1 & i \\ 
i & -1
\end{array}
\right) $
(from here on we take all constants to be real, unless otherwise
stated). Eqs. (\ref{eq:OF:B}-\ref{eq:OF:s}) are invariant under a
rotation by $\phi $ and a multiplication of $p$ by $e^{-i2\phi }$.

For the co-alignment function $a$ we take 
\begin{equation}
a\!\left( B,|p|\right) =-4D_\theta +\nu \,( B-|p|)/( B+|p|)
\end{equation}
The first term in the RHS results from linear diffusion of $\bd$ in
the $\theta$ dimension (with $D_\theta$ the diffusion coefficient) 
and the second term is an alignment of non-align bacteria with the
mean orientation (with $\nu$ being the rate of this co-alignment).
The exact form of $a$ is not important,
as long as $a\!\left(|p|\right)$ has at most one positive root
with negative derivative in the range $[0,B]$ and
$a\!\left(|p|\right)$ in non-positive outside the range $(0,B)$.

The rotation function $\gamma$ results from the only process that brakes
left-right symmetry, the bacterial bias in their tumbling when they are
placed in a thin layer of liquid.
The bacteria are restrained by their neighbors, and the restraints on
rotation are weakest near the boundary of the branches.
Due to the non-linear diffusion
\cite{GKCB98,Kitsunezaki97,Cohen97,KCGB99,CGKB99}, the branch's
boundary is identifiable by the divergence of $\nabla B$ (for
the initial conditions specified below).
The term $B^{k-1}\nabla B$ is always finite and can be used to
identify the boundary (as $p$ is bound by $B$, it can replace $B$ in
the coefficient).
$\gamma$ can be written as a function of $\Delta$, the angle 
between the branch's boundary and the bacterial mean
orientation: 
\begin{eqnarray}
\gamma \! \left( \vec{\alpha} ,\beta \right)
 &=& \,{\hbox{Re}}\!\left[ \gamma^0 \left| \beta \vec{\alpha} \right|
-\gamma^0 \beta  \left( {\vec{\alpha} ^T {\bf M} \vec{\alpha} }\right)
/ \left| \vec{\alpha} \right| \right]
\nonumber \\ &=& 
\left|\beta \vec{\alpha} \right| \left[
2{\hbox{Re}}\!\left(\gamma^0\right) \sin^2 \left(\Delta \right) +
{\hbox{Im}}\!\left(\gamma^0\right) \sin \left(2\Delta \right) \right] 
\end{eqnarray}
where $\gamma^0$ is a complex constant.
${\hbox{Re}}\!\left(\gamma^0\right)$ is a measure of the
anti-clockwise bias of the bacteria at the tip of the branch and
${\hbox{Im}}\!\left(\gamma^0\right) $ is a measure of the torque
aligning the bacteria in parallel to the boundary.
The rotation at the tip of the branch is also restrained by friction
with the agar, and from the
same reasoning that related $D_d$ and $D_l$ to the agar dryness,
we deduce that ${\hbox{Re}}\!\left(\gamma^0\right) $ is
inversely proportional to the agar dryness.

\begin{figure}[htbp]
\centerline{
 a
 \epsfxsize=1.2in
 \epsffile{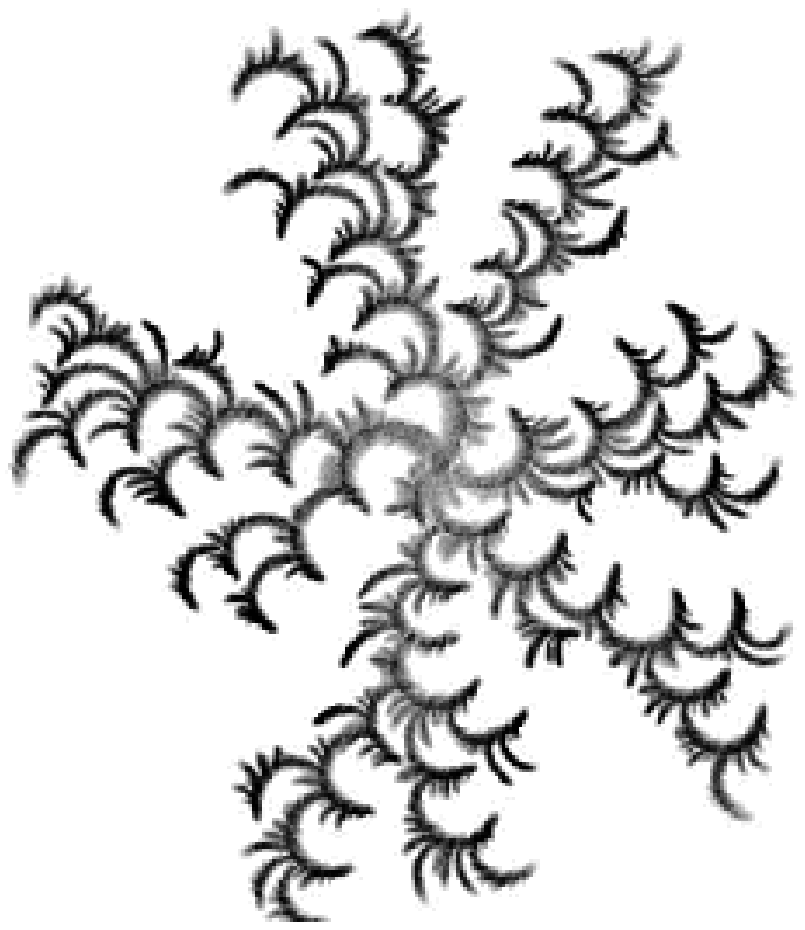}
 \ \ \ b
 \epsfxsize=1.2in
 \epsffile{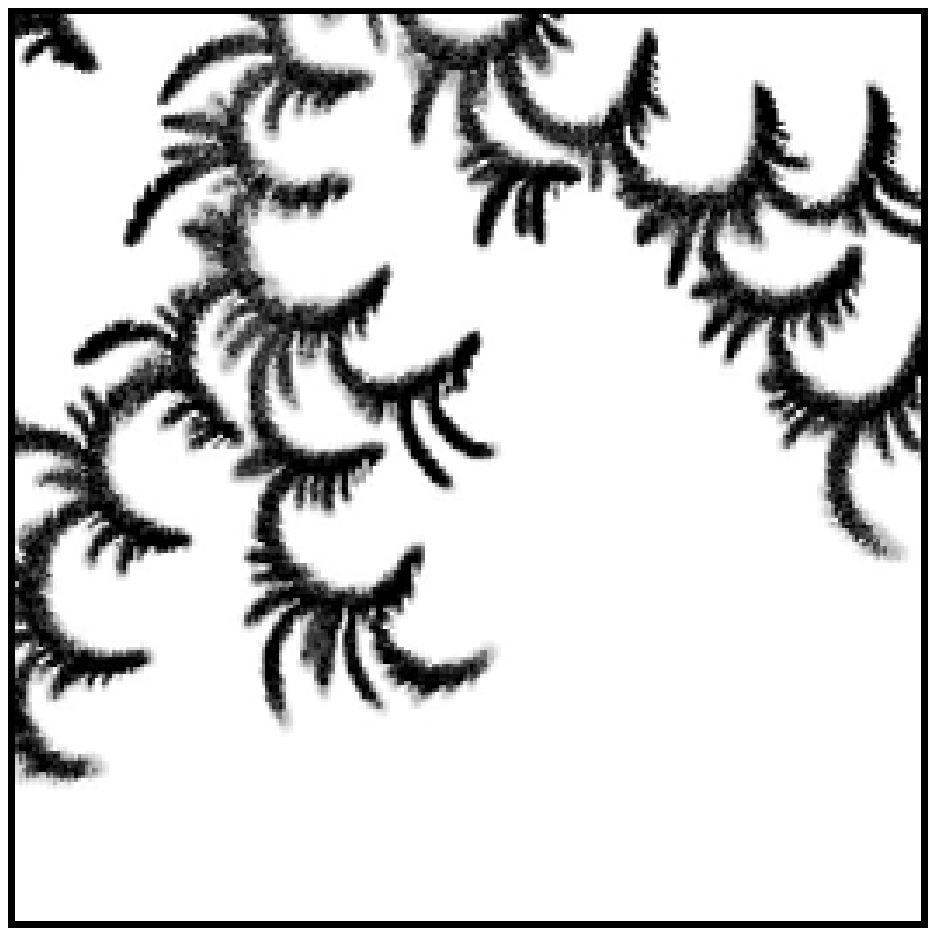}
}
\caption{
\label{fig:OF0}
Results of numerical simulation of the OF model.
a) Global view of a simulated colony with a local twist of branches.
Densities of $B+s$ are indicated by gray levels.
Parameters values are $k=3$, $D_d=0.0625$, $D_l=0$, $\mu=0.1$,
$D_\theta=0.001$, $\nu=0.5$, $\gamma^0=0.0075-i0.006$, $D_n=1$ and $n_0=1.5$.
b) A close look at the pattern of (a), showing the details of the
curved branches.
}
\end{figure}

As initial conditions (in a circular 2D geometry), we set $n$ to have
uniform distribution of level $n_0$, $B$ to have compact support at the
center where it is positive, and the other fields to be zero everywhere.
We solve the model numerically using a 2nd order explicit scheme.
In order to reduce the implicit anisotropy of the scheme, we use tridiagonal
lattice and multiply the bacterial diffusion coefficients by a quenched
noise with mean 1. The quenched noise represent inhomogeneities of the
agar surface. 
We show in Fig. \ref{fig:OF0} that the model can indeed reproduce the
microscopic bacterial dynamics and the chiral branching patterns with
the local twist.

The Non-Linear Diffusion model (NLD) for colonial growth can reproduce
tip-splitting branching patterns
\cite{GKCB98,Kitsunezaki97,Cohen97,KCGB99,CGKB99} of a related
morphotype, \Tme, with shorter bacterial cells.
The OF model reduces to the NLD model if the self-propulsion is not
primarily along the bacterial long axis ($D_d\simeq D_l$), if the
bacteria do not tend to co-align ($\nu \leq 4D_\theta $ or $\left( \nu
-4D_\theta \right) \ll \left( \nu +4D_\theta \right)$ ), or if
bacteria at the tip of the branch tend to rotate too freely
($\left|\gamma^0\right| \left( \nu -4D_\theta \right) > \left| {\bf
D_2} \right|$).
In all these cases $\left|{\hbox{Re}}\!\left( {\bf D_2} \nabla
p\right) \right| \ll \left| D_1 \nabla B \right|$ everywhere and the
OF model produce tip-splitting branching patterns.

\begin{figure}[htbp]
\centerline{
 a
 \epsfxsize=1.2in
 \epsffile{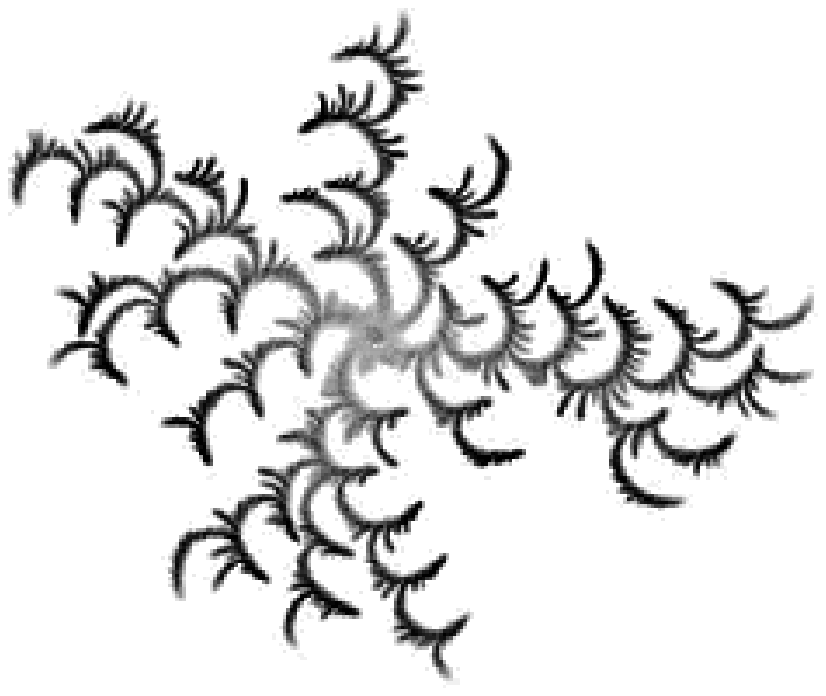}
 \ \ \ b
 \epsfxsize=1.2in
 \epsffile{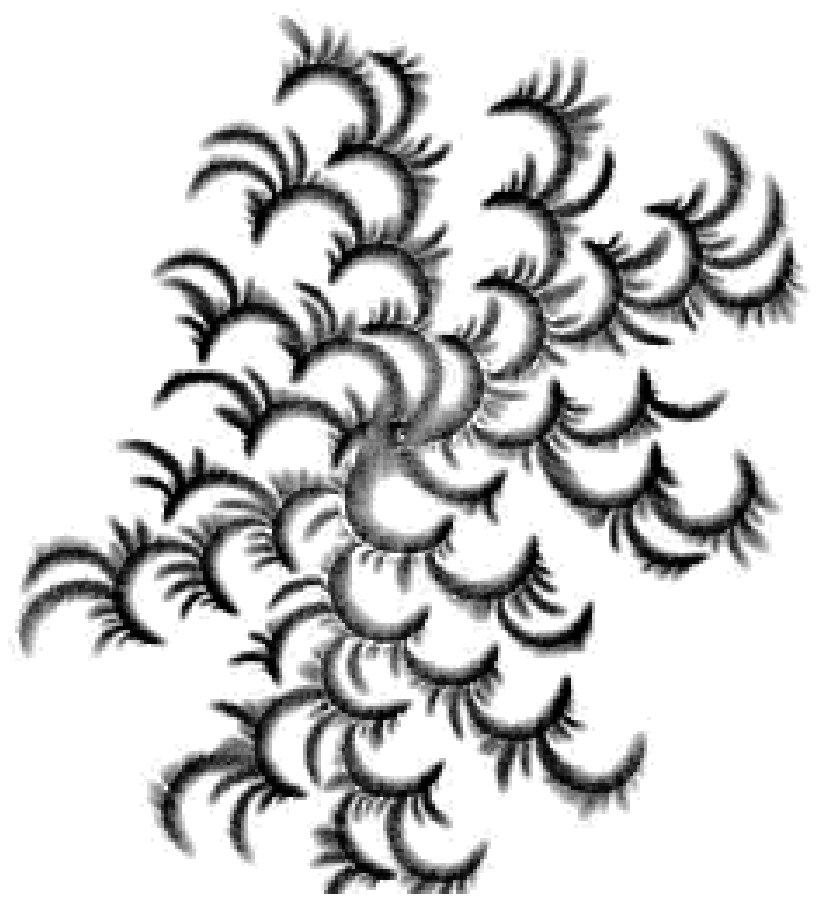}
}
\centerline{
 c
 \epsfxsize=1.2in
 \epsffile{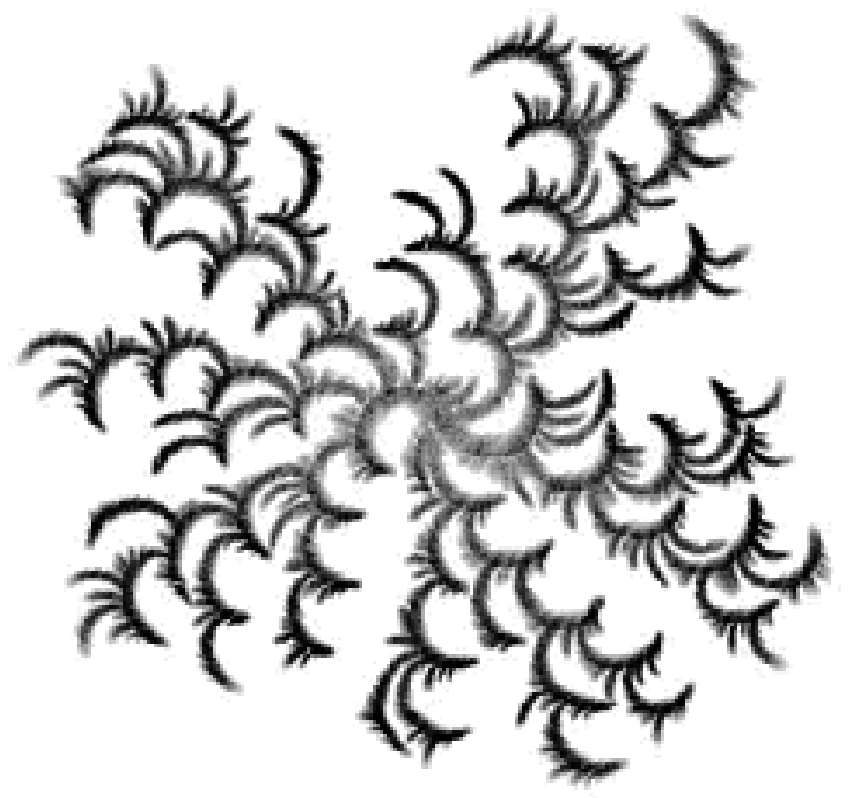}
 \ \ \ d
 \epsfxsize=1.2in
 \epsffile{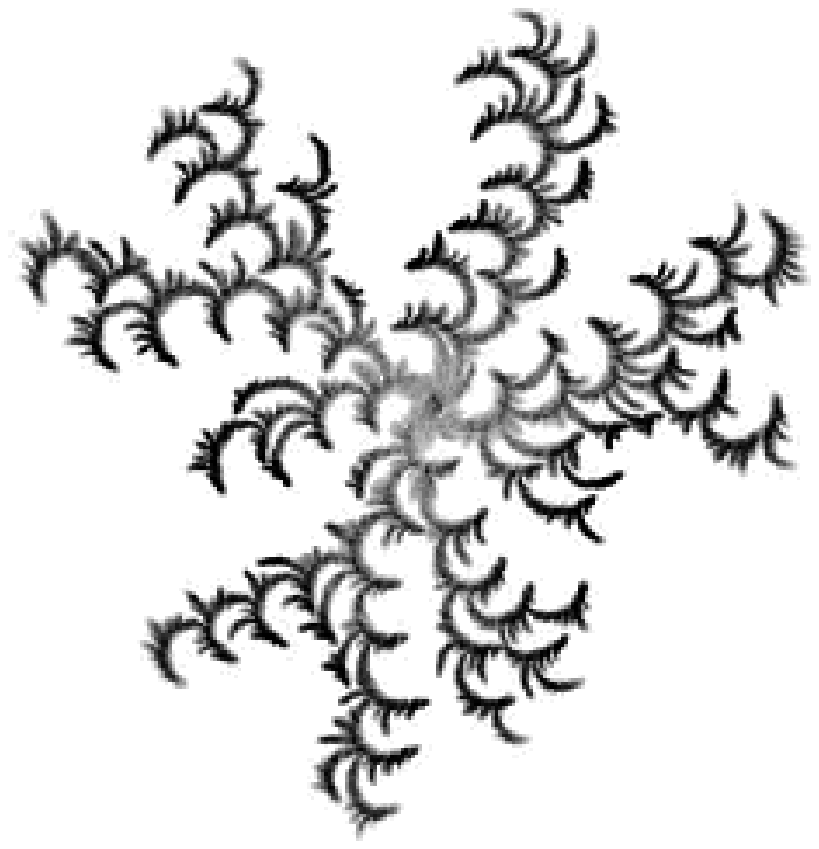}
}
\caption{
\label{fig:OF:md}
The response of simulated colonies to experimental control
parameters.
Figures (a) and (b) present change in initial food concentration with
$n_0=1$ and $n_0=2$ in (a) and (b) respectively. All other
parameters are the same as in Fig. \protect\ref{fig:OF0}a. The
expansion time of the colonies, normalized by the expansion time in Fig.
\protect\ref{fig:OF0}a, is 2.9 and 0.4 in (a) and (b) respectively.
Figures (c) and (d) present change in agar dryness $A$. $A$ is related
to the model's parameters with $\re\!\left(\gamma^0\right)=0.009/A$
and $D_d=0.09/A^2$. In (c) and (d) $A=1$ and $A=1.5$ respectively, with
all other parameters are the same as in Fig. \protect\ref{fig:OF0}a,
where $A=1.2$. The normalized expansion time of the colonies is 0.8 and
1.2 in  (c) and (d) respectively.
}
\end{figure}

The response of the simulated growth to initial food concentration and
agar dryness is shown in Fig. \ref{fig:OF:md}.
In agreement with experimental observations \cite{BCSCV95,BCGK99},
the main effect of the parameters is on
the global density of branches, while the curvature of the branches 
is only weakly affected.

\label{sec:chemo}

It was shown \cite{BSTCCV94a,BCSCV95,CCB96,BCGK99}
that many features of the colonial patterns are explained
when food chemotaxis and chemotactic signaling (a chemotactic response
to a material emitted by the bacteria themselves) are modeled.

In order to include chemotaxis in our model we bias the spatial
flux of $\bd$ in Eq. (\ref{eq:b:0}) by adding a chemotaxis term: 
$ B^k\bd\zeta_n (n) {\bf D_0}(\theta )\nabla n $ 
(here we take food chemotaxis as an example).
${\bf D_0}(\theta )\nabla n$
is the spatial derivative of the food concentration along the direction of
movement of the bacteria.
$\zeta_n (n) {\bf D_0}(\theta )\nabla n $ is the derivative as sensed
by the bacteria, where $\zeta_n (n)$ can be, for example, the
``receptor law'' \cite{Murray89} or a constant. The chemotaxis is
attractive for positive values of $ \zeta_n (n) $ and repulsive
otherwise.

The bacterial flux due to food chemotaxis translates in the OF model
into additional flux terms in Eqs. (\ref{eq:OF:B}-\ref{eq:OF:p}).
To the flux of $B$ in Eq. (\ref{eq:OF:B}) we add the term
$B^k\zeta_n (n) \left[B D_1 {\bf I} +2\,{\hbox{Re}}\!\left( p{\bf D_2}
\right) \right] \nabla n$ (where $ {\bf I}$ is the unit matrix)
and to the flux of $p$ in Eq. (\ref{eq:OF:p}) we add the term 
$B^k\zeta_n (n) \left[ {\bf D_2^{*}}B + p D_1 {\bf I}\right] \nabla
n$ .

If a repulsive signaling material $r$ is emitted by sporulating
cells \cite{BSTCCV94a,CCB96,BCGK99} then Eqs.
(\ref{eq:OF:B}-\ref{eq:OF:p}) are affected in a similar manner, with
$\zeta_r (r)$ replacing $\zeta_n (n)$ and $\nabla r$ replacing $\nabla
n$. $\zeta_r (r)$ is negative for a repulsive signal.
The equation for the dynamics of the signaling material is 
\begin{equation}
\dot{r}=D_r\nabla ^2r+\Gamma _rs-\lambda _BBr-\lambda _rr
\label{eq:OFc:r}
\end{equation}
where $\Gamma _r$, $\lambda _B$, $\lambda _r$ are non-negative constants. 
$\Gamma _r$ is the rate of chemical production by sporulating bacteria, 
$\lambda _B$ is the rate of chemical digestion by bacteria and 
$\lambda _r\,$ is the rate by which the chemical decompose.

\begin{figure}
\centerline{
 a
 \epsfxsize=1.2in
 \epsffile{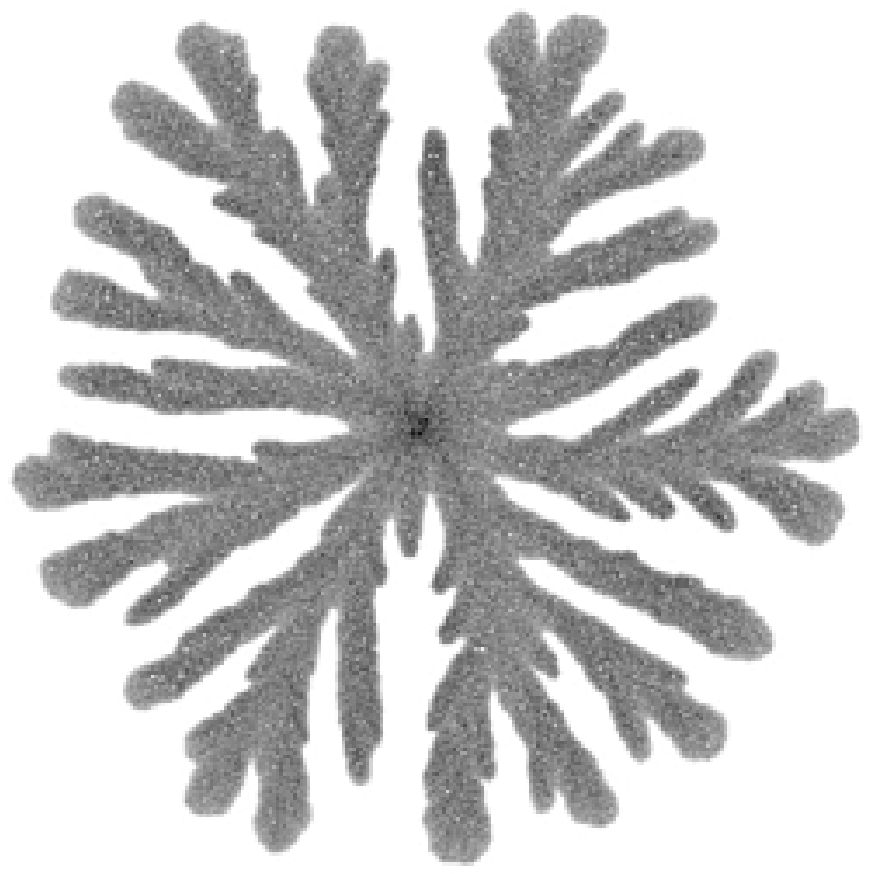}
 \ \ \ b
 \epsfxsize=1.2in
 \epsffile{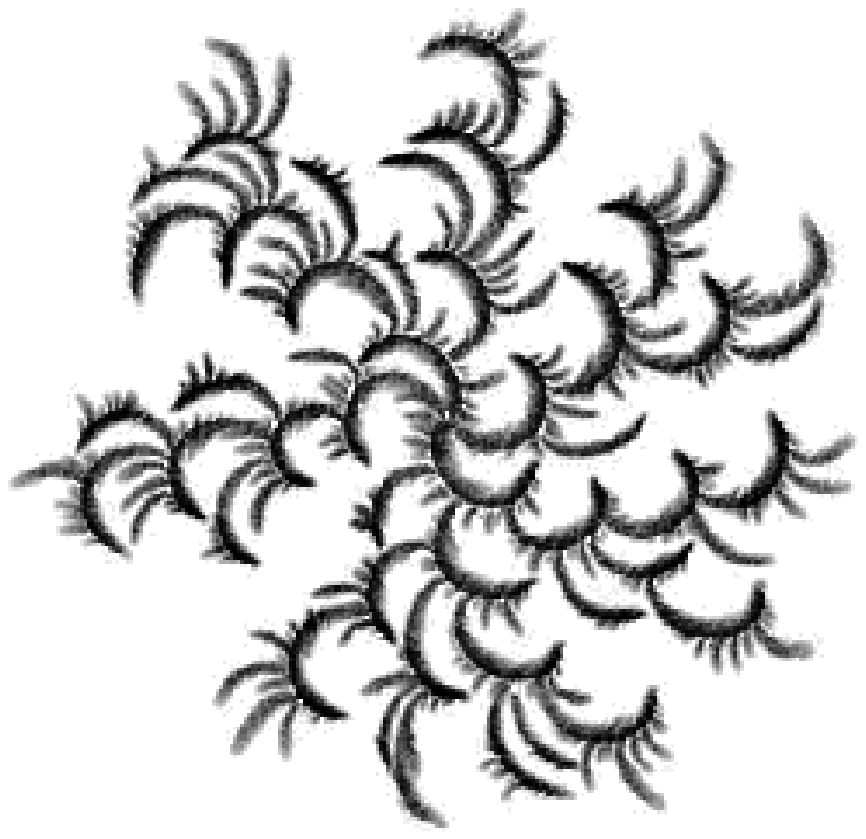}
}
\centerline{
 c
 \epsfxsize=1.2in
 \epsffile{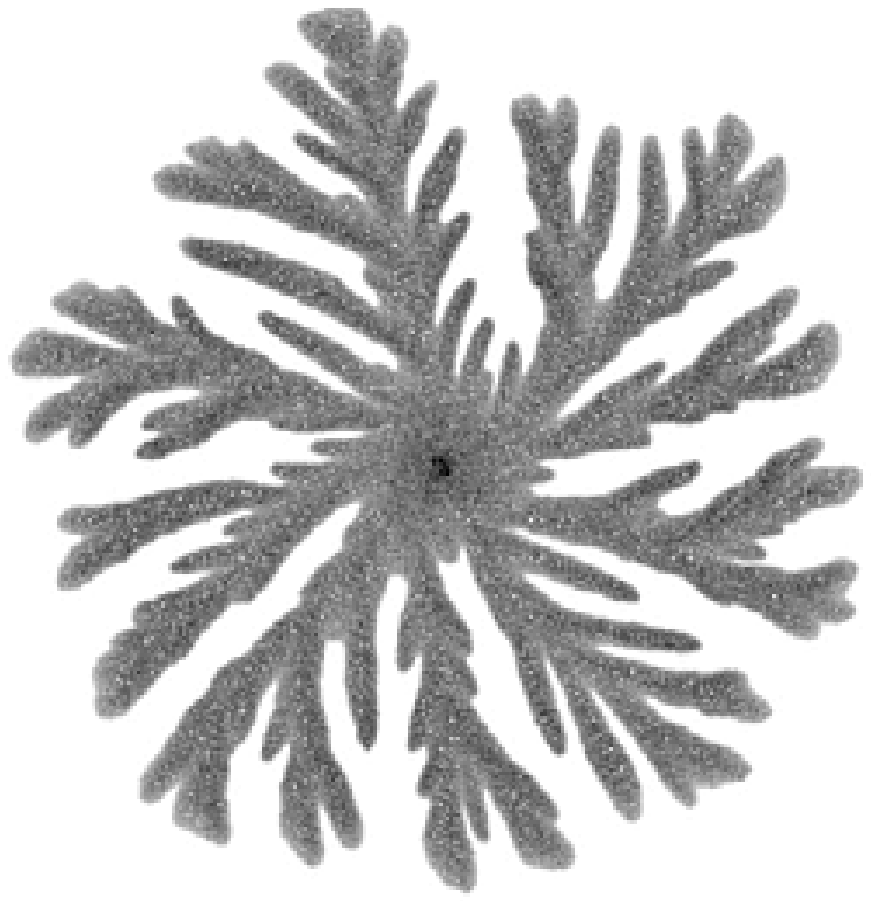}
 \ \ \ d
 \epsfxsize=1.2in
 \epsffile{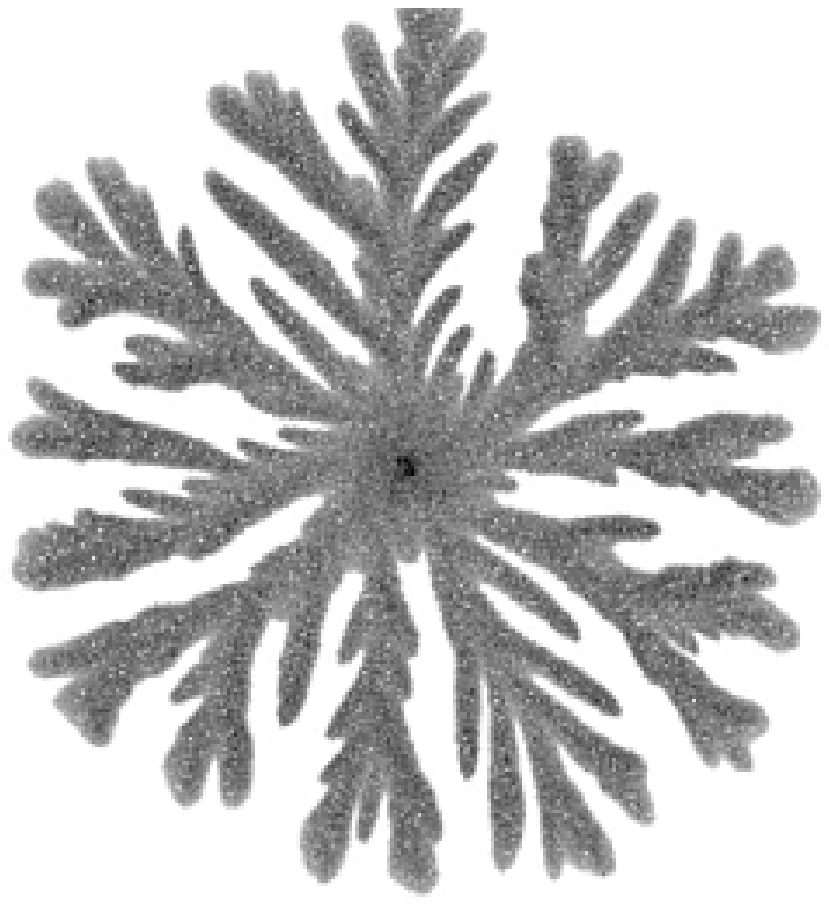}
}
\caption{
\label{fig:chemo}
a) The effect of repulsive chemotactic signaling on colonies
of \T bacteria.
To simulate short bacteria, unrestrained in rotation, we take
$D_\theta=0.256$ and $\re\!\left(\gamma^0\right)=0.128$.
For chemotaxis we take 
$\zeta_r(r)\equiv-1$,
$D_r=1.0$, $\Gamma_r=0.25$, $\lambda_B=0$ and $\lambda_r=0.01$.
All other parameters are as in \protect\ref{fig:OF:md}c.
Due to chemotaxis the branches are radially oriented with circular
global envelope.
b) The effect of repulsive chemotactic signaling on on colonies of
\Ce.
Most chemotaxis parameters are as in (b), $\Gamma_r=0.065$ and other
parameters are as in \protect\ref{fig:OF:md}c.
The branches acquire an outward bias and the global envelope becomes
circular.
c) Colony showing global twist in response to repulsive
chemotactic signaling. Parameters values are as in (b), with
$D_\theta=0.064$, $D_d=0.0625$ and $\re\!\left(\gamma^0\right)=0.096$,
representing bacteria of intermediate length on dryer surface.
d) ``De-chiraling'' the pattern in (c) by applying the mapping
$(r,\phi) \rightarrow (r,\phi + r/R )$, where $R$ is a constant, on
the polar coordinates $(r,\phi)$ measured from the center of the
colony.
}
\end{figure}

Both food chemotaxis and repulsive chemotactic signaling increase the
expansion rate of the colonies. Food chemotaxis does not affect the
colonial patterns significantly, while chemotactic signaling does
(Fig. \ref{fig:chemo}).
For short bacteria, the colonial pattern becomes less ramified, with
radial branches and circular global envelope. For long bacteria, the
global envelope becomes circular and the branches acquire an outward
bias, changing their appearance from an arc-like to a hook-like
appearance.

In Fig. \ref{fig:chemo} we also show the phenomena of global
twist \cite{BCSCV95,BCGK99}. For parameters representing bacteria
with intermediate length, repulsive chemotactic signaling can impose
global twist on an otherwise tip-splitting pattern.
The twist of the branches is relative to the center of the colony, not
to the local orientation of the branch. The global nature of the twist
is evident by using the ``de-chiraling'' method presented by
Ben-Jacob \etal in Ref. \cite{BCGK99} (see Fig. \ref{fig:chemo}d). 
In Ref.  \cite{BCGK99} Ben-Jacob \etal were able to obtain patterns
with a global twist, using the NLD model and
applying a rotation operator on the chemotaxis term. As
they argue themselves, such operator is inconsistent with the known
biological facts and a more detailed model -- like the OF model -- is
required in order to model any type of chirality in the bacterial
colonies.


In this letter we present a reaction-diffusion model which accounts
for the various morphologies presents by colonies of \dend-.
We focus on chiral features of the colonial patterns, but we are also
able to model tip-splitting patterns. We successfully simulated
intermediate growth patterns (not shown here).
The aim of developing such model is to aid in the study of transition
between the two types of growth.  Our model is able to handle events
of minute density such as a mutation in a single bacterium, and it
will be used in following studies of transitions between
morphologies and morphotypes.

We thank Inna Brainis for her technical assistance. Presented studies
are supported in part by a grant from the Israeli Academy of Sciences
grant no. 593/95.
IC thanks The Colton Scholarships for their support.

\end{multicols}
\widetext

\end{document}